\documentclass[aps,preprintnumbers,pacs,nofootinbib,secnumarabic]{revtex4}
\usepackage{graphicx}
\usepackage{amsmath}
\usepackage{amssymb}
\usepackage{xypic, mathtools}

\begin{document}

\title{Time vector defined in imaginary space of spatial coordinate}
\author{K. C. Wong$^{1}$}
\email{fankywong@gmail.com}
\author{P. L. Yu$^{1}$}
\email{yupuiling@gmail.com}

\affiliation{$^{1}$Department of Physics, University of Hong Kong, Pok Fu Lam Road, Hong
Kong, P. R. China}

\begin{abstract}
Empirical understanding teaches us that space is three dimensional while relativity merges space with time. We tried to show that it is possible to model space as three complex coordinates. In our construction, the usual spatial coordinate are the real part while time is considered as parameter of a path attached to each spatial point in imaginary parts. For flat spacetime, Lorentz invariant is realized on induced metric. We first consider the space as a six dimensional real manifold corresponding to the three complex coordinates. The complex structure is then implemented in a reference frame dependent manner. With this complex structure, Lorentz transformation induces a holomorphic transformation in the complex manifold so that spatial coordinate remained to be viewed as real axis. Scalar field in complex space could induce a field that satisfying Klein-Gordon equation on spacetime. Finally we described possible construction of gravity by using equivalence principle. Assuming the coordinate transformation to an accelerating observer also induces a holomorphic transformation, we outlined how to recover general relativity in Hamiltonian formalism.
\end{abstract}

\date{\today}
\maketitle

\section{Introduction}
The concept of spacetime originated from Einstein's theory of general relativity (GR). The theory has passed many observational tests of gravity, including perihelion precession, gravitational lensing, gravitational time dilation, corrections in global positing system, etc.

However there are numbers of problems associated with GR. On large scale it is very difficult to explain a small cosmological constant corresponding to the observed late time acceleration. Moreover, gravity may need to take account of dark matter problem in galactic and cluster scale. Modification of gravity models has been suggested in many aspects for these problems, some interesting examples are in \cite{Harko, Rham, Gia}.

On small scale GR face difficulty to be quantized as a field theory. In respond to this problem, recently it has been suggested that Lorentz symmetry could be an accidental symmetry of gravity \cite{Horava}. This construction has an indication that gravity can be quantized. On the other hand, \cite{Yu} demonstrated a reasonable model of gravity that only honors spatial diffeomorphism. The concept of time could be emerged from physical processes happened in space.

GR also presented a hierarchy problem in particle physics. Gravity is much weaker than all known particle interaction in all scale tested by human. Extra dimensional models has been proposed to explain this problem \cite{Nima, Randall}.

Time is a special coordinate axis, it is directional and irreversible. Therefore we look for an intrinsically different construction of time with respect to space. With three complex spatial coordinates, we are able to show that time could be parameters for a path in their imaginary parts. In order to recover special relativity, we think that the complex structure should be implemented after fixing a coordinate in spacetime. Coordinate transformation in spacetime that due to physical motions will be interpreted as holomorphic transformation in the corresponding complex structure. We demonstrate that it is possible for the case of Lorentz boost.

This paper is organized as follows. The second section is about coordinate system and special relativity (SR). We postulated the structure of space, and defined time by a path in imaginary parts of spatial coordinate and call it $\gamma$. We introduced a special matching of coordinate (a template) in complex space and real spacetime so that metric of complex space reproduces metric of special relativity. In the third section, the Lorentz transformation back and forth a template was lifted to holomorphic transformation in the complex space such that real axes remain to be spatial coordinate. Moreover, we considered how complex space change with spatial curvilinear coordinate transformation. In the fourth section, we showed that template frames do not necessarily lead to special frame of reference and constructed a scalar field in complex space that shows Lorentz invariant in spacetime. In fifth section, we discussed possible gravity model in this space and time structure. In the final section, we concluded our finding and discuss possible further researches.

We used Greek alphabet to denote tensor index in 4D, $\mu=0,1,2,3$. Small roman letter for tensor index in 3D $i=1,2,3$ and capital roman are for tensor index in complex space $M=\tilde{1},\tilde{2},\tilde{3},1,2,3$ where $\tilde{i}$ is the coordinate associated with imaginary parts of $x^i$. We also use subscript in time to label instead of indicating component. For example, $t_x$ mean time in coordinate associated with $x^i$. We use $\Lambda^{\mu'}_{\nu}$ to denote Lorentz transformation $\partial_{\nu} x'^{\mu}$ from $x_{\mu}$ to $x'_{\mu}$ with $\partial_{\nu}$ denote derivative with respect to $x^{\nu}$. The imaginary number $\sqrt{-1}$ is denoted by $i$ and there should be no confusion with the tensor index.
\section{Complex coordinate and special relativity}
We started with a six dimensional real smooth manifold $\mathcal{C}$ described by three complex coordinates. Let the coordinates be
\begin{eqnarray}\label{XYZ}
X^1=x^1+ic^1,\\
X^2=x^2+ic^2,\\
X^3=x^3+ic^3.
\end{eqnarray}
Let's call our usual description of the world as 4D Lorentzian manifold $\mathcal{M}^4$, and fix a coordinate $(x^i,t_x)$. If physics in $\mathcal{M}^4$ can be realized in $\mathcal{C}$, there should be an embedding $\gamma:\mathcal{M}^4\mapsto\mathcal{C}$ such that everything happening in our world can be put into complex space. We are looking for a special type
\begin{equation}\label{lessgood}
\gamma^i(x,t_x)=x^i+ic^i(x,t_x),
\end{equation}
so that time vector is a vector in imaginary space (space spanned by $\partial_{c^i}$)
\begin{equation}\label{timevector}
\frac{\partial}{\partial t}=\frac{\partial c^i}{\partial t}\frac{\partial}{\partial{c^i}}.
\end{equation}
This embedding allows us the treat time as parameter of a path in imaginary space attached to each spatial point. Moreover, $\gamma$ pullback the metric in $\mathcal{C}$ to $\mathcal{M}^4$. $\mathcal{C}$ should has a metric $\tau$ in coordinate $X^i, \bar{X}^i$, e.g.
\begin{equation}\label{flatcomplex}
\tau=\frac{dX^1dX^1+d\bar{X}^1d\bar{X}^1}{2}+\frac{dX^2dX^2+d\bar{X}^2d\bar{X}^2}{2}+\frac{dX^3dX^3+d\bar{X}^3d\bar{X}^3}{2},
\end{equation}
which could give a metric in $(x^i,c^i)$ coordinate
\begin{equation}\label{flatcomplexR}
{}^{(6)}ds^2=-\left(dc^1dc^1+dc^2dc^2+dc^3dc^3\right)+\left(dx^1dx^1+dx^2dx^2+dx^3dx^3\right).
\end{equation}
In this case, if we can set an embedding of
\begin{equation}\label{preferE}
\gamma_x(t_x, x^i)=(x^i,c_x^1(t),c_x^2(t),c_x^3(t)),
\end{equation}
$\gamma$ will pullback the metric in $\mathcal{C}$ to a metric
\begin{equation}\label{time}
-\left[\sum_{i=1}^{3}(\dot{c}_x^i)^2\right]dt^2+\left(dx^1dx^1+dx^2dx^2+dx^3dx^3\right)
\end{equation}
which is in $\mathcal{M}^4$. For embedding that satisfy
\begin{equation}\label{normal}
\left[\sum_{i=1}^{3}(\dot{c}_x^i)^2\right]=1
\end{equation}
for all points $p$, the metric in SR
\begin{equation}\label{metricSR}
ds^2=-dt^2+dx^1dx^1+dx^2dx^2+dx^3dx^3
\end{equation}
can be reproduced. In this section, we used $x^i,t_x$ in $\mathcal{M}^4$ and $X^i$ in $\mathcal{C}$ to denote coordinates that realized Eq.~(\ref{flatcomplex}) Eq.~(\ref{preferE}) and Eq.~(\ref{normal}) with constant $\dot{c}^i$ so that special relativity is recovered. We called this a template frame.
\section{Change of coordinate on template}
\subsection{Holomorphic map and Lorentz transformation}\label{lor}
The template frame allows us to recover metric of SR on $\mathcal{M}^4$ which is Lorentz invariant. Suppose there is a Lorentz transformation $\Lambda^{\mu '}_{\nu}$ in $M^4$ such that it transforms $(x^i,t_x)$ into $(y^i,t_y)$. Let's call the corresponding embedding to be $\gamma_x$ and $\gamma_y$. There should be a coordinate transformation in $\mathcal{C}$ such that $\gamma_y$ takes the form of Eq.~(\ref{lessgood}), i.e. a holomorphic transformation $X^i\to Y^i$ such that $y^i=Re(Y^i)$. This means that $Y^i$ has $y^i$ as real parts, and time can still be viewed as a vector in imaginary space, i.e. Eq.~(\ref{timevector}), for observers in $(y^i,t_y)$. Therefore we defined lifting of $\Lambda$ to be a holomorphic transformation in $\mathcal{C}$, written as $\Phi+i\Psi$ with $\Phi$ and $\Psi$ being the real and imaginary part respectively, such that the following diagram commutes, i.e. $\gamma_y(\Lambda(x^{\mu}))=\Phi(\gamma_x(x^{\mu}))+i\Psi(\gamma_x(x^{\mu}))$.
\begin{equation*}
    \xymatrix{
        *+[o+2][F]{\mathcal{C}_Y, Y^i} \ar @{<->}[r] ^{\Phi+i\Psi} & *+[o+2][F]{\mathcal{C}_X, X^i} \\
        *+[o+2][F]{M_y, (y^i,t_y)} \ar[u]^{\gamma_y}  \ar @{<->} [r]_{\Lambda} & *+[o+2][F]{M_x, (x^i,t_x)} \ar[u]_{\gamma_x} }
\end{equation*}

On one hand we can push $\Lambda$ to a map $M_x\mapsto\mathcal{C}_Y$ by $\gamma_y$,
\begin{equation}
y^i(x,t_x)+ic_y^i(y(x,t_x),t_y(x,t_x)).
\end{equation}
Differentiating it we obtain
\begin{eqnarray}\label{set1}
\frac{\partial y^i}{\partial x^j}=\Lambda^{i '}_{j},&&\frac{\partial y^i}{\partial t_x}=\Lambda^{i'}_{0},\\
\frac{\partial c_y^i}{\partial x^j}=\partial_{\mu}c_y^i\Lambda^{\mu '}_{j},&&\frac{\partial c_y^i}{\partial t_x}=\partial_{\mu}c_y^i\Lambda^{\mu '}_{0}.
\end{eqnarray}

On the other hand we can map $M_x$ to $\mathcal{C}_Y$ by using $\gamma_x$ followed by a holomorphic transform in $\mathcal{C}$
\begin{eqnarray}
\Psi^i(x,c_x(t_x))&=&c_y^i,\\
\Phi^i(x,c_x(t_x))&=&y^i.
\end{eqnarray}
From this we can calculate
\begin{eqnarray}\label{set2}
\frac{\partial y^i}{\partial x^j}=\frac{\partial\Phi^i}{\partial x^j},&&\frac{\partial y^i}{\partial t_x}=\frac{\partial\Phi^i}{\partial c_x^j}\dot{c}_x^j,\label{set2a}\\
\frac{\partial c_y^i}{\partial x^j}=\frac{\partial\Psi^i}{\partial x^j},&&\frac{\partial c_y^i}{\partial t_x}=\frac{\partial\Psi^i}{\partial c_x^j}\dot{c}_x^j.\label{set2b}
\end{eqnarray}
If $\Phi+i\Psi$ is a holomorphic transformation, it satisfies Cauchy-Riemann (CR) equations
\begin{eqnarray}
\frac{\partial \Phi}{\partial x^j}&=&\frac{\partial \Psi}{\partial c_x^j},\\
\frac{\partial \Psi}{\partial x^j}&=&-\frac{\partial \Phi}{\partial c_x^j}.
\end{eqnarray}
Comparing with Eq.~(\ref{set2}) and eliminating $\Phi$ we obtain constraint equations on $\gamma_y$.
\begin{eqnarray}
\Lambda^{i '}_{j}\dot{c}_x^j=\Lambda^{\mu '}_{0}\partial_{\mu}c_y^i,\label{con1}\\
-\partial_{\mu}c_y^i\Lambda^{\mu '}_{j}\dot{c}_x^j=\Lambda^{i'}_{0}.\label{con2}
\end{eqnarray}
We should find $\partial_{\mu}c_y^i$ for a given $\Lambda^{\mu'}_{\nu}$ and $\dot{c}_x^i$.

If $\Lambda$ is a boost, $\Lambda^{i'}_0=-\Gamma v^i$ where ${\Gamma}$ is the Lorentz factor. If we choose, e.g.
\begin{equation}
\partial_0c_y^i=\dot{c}_x^i.
\end{equation}
Then Eqs.~(\ref{con1},\ref{con2}) become problem of finding a matrix $\partial_j c^i_y$ with images of two vectors specified, i.e.
\begin{eqnarray}
\partial_j c^i_y(\Lambda^{j'}_0)=\Lambda^{i '}_{k}\dot{c}_x^k-\Lambda^{0'}_0\dot{c}_x^i,\label{con3}\\
\partial_j c^i_y(\Lambda^{j'}_k\dot{c}^k_x)=-\Lambda^{0'}_k\dot{c}^k_x\dot{c}^i_x-\Lambda^{i'}_0.\label{con4}
\end{eqnarray}
If $\Lambda^{j'}_k\dot{c}^k_x=\Lambda^{j'}_0$, these two equations may be inconsistent so that lifting is not possible. However, if this happens we can use the explicit form of boost $\Lambda^{i'}_j=(\Gamma-1)v^iv_j/v^2+\delta^i_j$ to show that $\dot{c}^i_x$ are proportional to $v^i$
\begin{equation}
\dot{c}^i_x=\left[\Gamma-\frac{(\Gamma-1)v_j\dot{c}_x^j}{v^2}\right]v^i,
\end{equation}
this would imply $v=1$ by contracting $v_i$ on both sides. The velocity limited by Lorentz transformation ($v<1$) guarantee a solution of $\partial_j c^i_y$. After solving the $\partial_j c^i_y$ which are constants, functions $c^i(t_y,y)$ can then be taken as linear transformation of $t_y$ and $y$.

For spatial rotation, $\Lambda^{0'}_0=1$ and $\Lambda^{i'}_0=0$. Now we can pick $\partial_jc_y^i=0$ so that Eq.~(\ref{con2}) vanishes, and Eq.~(\ref{con1}) gives
\begin{equation}
\Lambda^{i'}_{j}\dot{c}^j_x=\dot{c}^j_y,
\end{equation}
which said that $\dot{c}_x$ and $\dot{c}_y$ is related by a rotation $\Lambda^{i'}_{j}$. Spatial rotations can be lifted to holomorphic transformation despite the fact that it is not corresponding to a physical motion.
\subsection{Inverse transform to the template}
Suppose $\Phi+i\Psi$ is invertible, the inverse could serve as a lift of $\Lambda^{-1}$. Therefore we need to show that the holomorphic map satisfying
\begin{eqnarray}
\frac{\partial \Phi^i}{\partial x^j}&=&\frac{\partial \Psi^i}{\partial c^j}=\Lambda^{i'}_j,\\
\frac{\partial \Phi^i}{\partial c^j}\dot{c}^j_x&=&-\frac{\partial \Psi^i}{\partial x^j}\dot{c}^j_x=\Lambda^{i'}_0,
\end{eqnarray}
could have an inverse. Let's denote the real Jacobian matrix as $J$. For $\Phi+i\Psi$ to have a local holomorphic inverse, we need $\det(J)\neq0$. If we choose $\Phi$ and $\Psi$ to be linear functions of $(x^i,c^i)$, non-zero determinant implies that the map is globally invertible.

For boost we can pick
\begin{equation}
\frac{\partial \Psi^i}{\partial x^j}=H^i_j=\left\{
\begin{array}{ll}
\Gamma v^i/a\dot{c}^j, &\mbox{for $\dot{c}^j_x\neq0$},\\
0, &\mbox{for $\dot{c}^j_x=0$},
\end{array}
\right.
\end{equation}
where $a$ is the number of non-zero component in $\dot{c}^j$. The Jacobian matrix in ordered basis $(x^i,c^i)$ is
\begin{equation}
J=
\begin{pmatrix}
\Lambda^{i'}_j& -H^i_j\\
H^i_j& \Lambda^{i'}_j
\end{pmatrix}.
\end{equation}
The determinant can be computed by block
\begin{equation}
\det(\Lambda^{i'}_j)\det(\Lambda^{i'}_j+H(\Lambda^{i'}_j)^{-1}H),
\end{equation}
which can be calculated to be
\begin{equation}
\det(J)=\Gamma^2+(H^i_i)^2.
\end{equation}
Because the first term is strictly positive and second term is positive, the determinant will also be non-zero.

For spatial rotation, we have $H=0$ and the determinant will also be non-zero.
\subsection{Complex spatial axis in curvilinear coordinates}\label{coordinateTrans}
We would like to see how change of spatial coordinates work in the complex space. For example, if we use spherical coordinate $(r,\theta,\phi)$, can we just add imaginary parts to all coordinates?

Suppose we start with a spatial coordinate transformation from $x^i$ to $u^i$, for example,
\begin{equation}\label{uvw}
u^i=u^i(x^1,x^2,x^3).\\
\end{equation}
We look for a lift of $u^i$ to $U^i=\Phi^i+i\Psi^i$ such that $\Phi^i=u^i$. This implies that
\begin{equation}
\frac{\partial \Phi^i}{\partial c^j}=0
\end{equation}
for all $i$ and $j$, i.e. $\Psi$ can have no dependency on $x^i$ by CR equations. Therefore $U^i$ can only be the form
\begin{equation}
U^i=u^i(x^1,x^2,x^3)+ic_u^i(c_x^1,c_x^2,c_x^3).
\end{equation}
Another CR equation
\begin{equation}
\frac{\partial u^i}{\partial x^j}=\frac{\partial c_u^i}{\partial c_x^j},
\end{equation}
can be generally true only if they are constants. Therefore spatial curvilinear coordinate transformations do not generally lift to holomorphic transformations. The main reason that it cannot be lifted because $\Phi$ has no dependency of $c_x^i$, which is true for transformations that do not involve time. However $U^i$ is a smooth coordinate transformation if $\mathcal{C}$ is considered as a 6D real manifold.
\section{Complex structure from template}
\subsection{No special frame of reference}
We remark that Eq.~(\ref{con1},\ref{con2}) do not allow $\partial_j c^i_y=0$. Therefore it is impossible to have a non-trivial holomorphic transformation in $\mathcal{C}$ such that Lorentz transformation of a template frame remains to be a template.

The consequence is that each set of inertial reference frames have only one template. However, this does not necessary mean that template frame is physically special.

From section (\ref{lor}), we showed that it is always possible to find a holomorphic coordinate in $\mathcal{C}$ such that spacetime is embedded like Eq.~(\ref{lessgood}) with $c^i(x,t_x)$ linear transforms of $(x,t_x)$. If $c^i(x,t_x)$ is linear, the smooth structure of $\mathcal{C}$ allows us to change any reference frames into a template. However a change of coordinate in $\mathcal{C}$ that lead to a change of template is not holomorphic.

Suppose we use $X^i$ in $\mathcal{C}$ to realize $(x,t_x)$ as template, the collection of all charts in $\mathcal{C}$ (as a real manifold) that are holomorphic to $X^i$ could define a holomorphic atlas. This will allow us to implement the complex structure, which is template frame dependent. On the other hand, we think that the template frame is a tool to visualize the complex structure of $\mathcal{C}$ in a reference frame dependent manner.

We use a scalar field as an example to see how 4D physics can deduce no special frame of reference.
\subsection{Scalar fields in $\mathcal{C}$}
A scalar field $\phi(X)$ on $\mathcal{C}$ with Lagrangian $\mathcal{L}(\phi, \partial_{c^i}\phi, \partial_i\phi)$ pullback to scalar field $\phi^*=\phi(\gamma_x(X))$ on $\mathcal{M}^4$. However, the Lagrangian cannot pullback from $\mathcal{C}$ therefore the dynamics of $\phi$ have to be determined in $\mathcal{C}$. For example,
\begin{equation}
\mathcal{L}=\frac{1}{2}\left(\tau^{MN}\partial_M\phi\partial_N\phi\right)-V(\phi)
\end{equation}
will lead to dynamics
\begin{equation}
-\partial_M(\tau^{MN})\partial_N\phi-\frac{1}{\sqrt{-\tau}}\tau^{MN}\partial_M(\sqrt{-\tau})\partial_N\phi
-\tau^{MN}\partial_M\phi\partial_N\phi+V'(\phi)=0.
\end{equation}
Since $\tau^{MN}=\gamma^M_{,\mu}\gamma^N_{,\nu}\eta^{\mu\nu}$ for points in $\mathcal{M}^4$, the dynamics of $\phi$ will agree with $\phi^*$ which satisfying
\begin{equation}
\square\phi^*-\partial^{\mu}(\ln\sqrt{-\tau})\partial_{\nu}\phi^*-V'(\phi^*)=S,
\end{equation}
where $S$ is obtained by evaluating $\partial_M(\tau^{MN})\partial_N\phi$ on $\mathcal{M}^4$. If $\tau$ is induced by a 4D Lorentz transformation on template frame, it is a constant. This can deduced by comparing Eq.~(\ref{set1}) with Eq.~(\ref{set2}) and making use of CR equations. Therefore $\phi^*$ agree with Klein-Gordon equation in flat space
\begin{equation}
\square\phi^*-V'(\phi^*)=0,
\end{equation}
which is Lorentz invariant. There is no different whether $\phi^*$ is evaluated on template frame or not.
\section{Gravitation}
\subsection{The equivalence principle and physical change of coordinates}
In section (\ref{coordinateTrans}) we found that general spatial coordinate transformation do not lift to holomorphic transformation. However, a change of coordinates that corresponds to physical motion must involve time. If the lifting of $\Lambda$ can be generalized to transformation that corresponding to accelerated motion, we can understand gravity in the setting of complex space by using equivalent principle. Suppose we have a physical coordinate transform (coordinate transform due to physical motion) from $(x^i,t_x)$ to $(y^i,t_y)$, denoted by $J^{\mu'}_{\nu}$. Assume $(x^i,t_x)$ to be free falling observer who find space and time being locally flat, $(y^i,t_y)$ will find gravity. The transformation $J^{\mu'}_{\nu}$ depends on both space and time. We go back to the constraint equations that allow a lifting, analogous to Eq.~(\ref{con1}) and Eq.~(\ref{con2})
\begin{eqnarray}
J^{i '}_{j}\dot{c}_x^j=J^{\mu '}_{0}\partial_{\mu}c_y^i,\label{con1g}\\
-\partial_{\mu}c_y^iJ^{\mu '}_{j}\dot{c}_x^j=J^{i'}_{0},\label{con2g}
\end{eqnarray}
We also need Eq.~(\ref{set2a}) and Eq.~(\ref{set2b}) to find the explicit form of $\Phi$ and $\Psi$. We cannot find a way to integrate the system in a general case. However if we assume this system can be integrated, we can investigate how gravity related to geometry in complex space $\mathcal{C}$.
\subsection{Curvature}
Assume that above equations can be integrated, there are constraints on $\tau$ because it must be related to Eq.~(\ref{flatcomplex}) by holomorphic transformation and it must pullback to a real symmetric metric by $\gamma_y$. According to equivalent principle, they are related by a physical transformation in $\mathcal{M}^4$.

In real coordinate $(x^i,c_x^i)$, a holomorphic transform to $(y^i,c_y^i)$ define two transformation matrices
\begin{equation}
{\bf j}=\frac{\partial y^i}{\partial c_x^j} \mbox{ and } {\bf f}=\frac{\partial y^i}{\partial x^j}
\end{equation}
Being holomorphic to Eq.~(\ref{flatcomplexR}) implies that $\tau$ is of the form

\begin{eqnarray}
\tau&=&
\begin{pmatrix}
{\bf j}& -{\bf f}\\
{\bf f}& {\bf j}
\end{pmatrix}
\begin{pmatrix}
{\bf -1}_{3\times 3}& {\bf 0}\\
{\bf 0}& {\bf 1}_{3\times 3}
\end{pmatrix}
\begin{pmatrix}
{\bf j}& -{\bf f}\\
{\bf f}& {\bf j}
\end{pmatrix}\\
&=&
\begin{pmatrix}
-{\bf u}& {\bf v}\\
{\bf v}& {\bf u}
\end{pmatrix},
\end{eqnarray}
where ${\bf u}=({\bf jj}+{\bf ff})$ and ${\bf v}={\bf jf}-{\bf fj}$. $\gamma_y$ will pullback $\tau$ into 4D metric $g_{\mu\nu}$ by
\begin{equation}
\tau_{MN}\gamma^M_{,\mu}\gamma^N_{,\nu}=g_{\mu\nu},
\end{equation}
we have omitted subscript $y$ in $\gamma$ for simplicity and used comma for ordinary derivative.
\subsection{Comparing with general relativity}
In order to recover GR, we use a $3+1$ formulation of GR on time slicing,
\begin{equation}
ds^2=h_{ij}(dx^i+N^idt)(dx^j+N^jdt)-N^2dt^2
\end{equation}
Identifying 4D metric function with metric derived from $\tau$, we find
\begin{enumerate}
\item{The shift vector $N_i$ is $-{\bf u}_{mk}c^m_{,i}\dot{c}^k+{\bf v}_{ik}\dot{c}^k$;}
\item{Spatial metric $h_{ij}$ is ${\bf u}_{ij}+{\bf v}_{mk}c^m_{,i}c^k_{,j}$;}
\item{The lapse $N$ can be calculated from $N^jN_j-N^2={\bf u}_{mk}\dot{c}^m\dot{c}^k$.}
\end{enumerate}
We used dot to denote time derivative. These variables follow Hamiltonian of Arnowitt-Deser-Misner formulation of GR \cite{ADM}, at least correct up to tested precision of gravity. Despite of lacking theoretical beauty, the complex metric $\tau$ and the embedded coordinate $c_y^i$ could reproduce gravity.
\section{Conclusion and Discussion}
\subsection{Conclusion}
We have demonstrated that in flat spacetime it is possible for space to be complex algebraic field with time defined by a path in imaginary space attracted to each spatial point. In this setting we give an template frame with coordinate $(x^i,c^i)$ in $\mathcal{C}$ and coordinate in $(x^i,t_x)$ in $\mathcal{M}^4$ to recover the metric in SR. The space is considered as six dimensional real manifold that has a complex structure after fixing a template.

Then we have shown that Lorentz transformation back and forth the template can be lifted to a holomorphic transformation under the requirement that real axis in $\mathcal{C}$ align with spatial coordinate in $\mathcal{M}^4$, i.e. Eq.~(\ref{lessgood}). In general the holomorphic transformation will change the metric of complex space $\tau$ and Lorentz symmetry in $\mathcal{M}^4$ is recovered by the embedding. We also demonstrated that general curvilinear spatial coordinate transformation do not lifted to holomorphic transformation.

The template frame is fixed by smooth structure of $\mathcal{C}$, therefore we are not embedding the whole spacetime into a three dimensional complex manifold. The complex structure is implemented after fixing a reference frame in $\mathcal{M}^4$ with Lorentz transformation honoring this complex structure. Therefore it is a preferred reference frame in $\mathcal{C}$ with respect to a reference frame in $\mathcal{M}^4$. We used a real scalar field in $\mathcal{C}$ as an example for Lorentz invariant physics in $\mathcal{M}^4$.

With the assumption that reference frame of accelerating observer can be lifted to holomorphic transform, we investigated how complex space is related to gravity. However, whether a gravity theory can be constructed in complex space like scalar field is non-trivial, it is left for further study.

\acknowledgments

We would like to thank Dr.~Tiberiu Csaba Harko, Mr.~Yu Hoi Fung and Miss.~Wong Ching Yat in the University of Hong Kong for important discussions and suggestions about this work.

\end{document}